\documentclass{scrartcl}
\usepackage{graphicx}
\usepackage{dcolumn}
\usepackage{bm}

\usepackage[english]{babel}
\usepackage[T1]{fontenc}
\usepackage[utf8]{inputenc}
\usepackage{amsmath}
\usepackage{amssymb}
\usepackage[thinspace,mediumqspace,amssymb]{SIunits}
\begin{document}

\title{Phonon band structures of three-dimensional pentamode metamaterials}

\author{Aude Martin$^1$, Muamer Kadic$^1$, Robert Schittny$^1$,\\ Tiemo Bückmann$^1$, and Martin Wegener$^{1,2}$}
\maketitle

\begin{flushleft}
\it{
$^1$Institute of Applied Physics and DFG-Center for Functional Nanostructures (CFN), Karlsruhe Institute of Technology (KIT), 76128 Karlsruhe, Germany\\
$^2$Institute of Nanotechnology, Karlsruhe Institute of Technology (KIT), 76128 Karlsruhe, Germany}
\end{flushleft}

\begin{abstract}
Three-dimensional pentamode metamaterials are artificial solids that approximately behave like liquids, which have vanishing shear modulus. Pentamodes have recently become experimental reality. Here, we calculate their phonon band structures for various parameters. Consistent with static continuum mechanics, we find that compression and shear waves exhibit phase velocities that can realistically be different by more than one order of magnitude. Interestingly, we also find frequency intervals with more than two octaves bandwidth in which pure single-mode behavior is obtained. Herein, exclusively compression waves exist due to a complete three-dimensional band gap for shear waves and, hence, no coupling to shear modes is possible. Such single-mode behavior might, e.g., be interesting for transformation-elastodynamics architectures.
\end{abstract}

   
\maketitle

\nocite{*}
Artificial materials called metamaterials have recently started to emerge in acoustics and elastodynamics, for example see \cite{Cummer08,Zhang11,Lai11,Stenger12,Liang12,Christensen12,Kadic12,Torrent12,Mei12}. This field is driven by the quest for more control on elastic/acoustic waves, inspired by corresponding progress on three-dimensional metamaterials in electrodynamics/optics (for recent reviews see \cite{Soukoulis11,Moser12}). Three-dimensional (3D) pentamode metamaterials can be seen as the elastodynamic counterpart of 3D magnetodielectric metamaterials in optics in the following sense: In 3D transformation optics \cite{Pendry06,Leonhardt10}, magnetodielectrics enable anisotropies without birefringence and, hence, an optical wave impinging onto some structure with a certain polarization also emerges with that polarization and does not generate partial waves with other polarizations. In mechanics or, more precisely, in elastodynamics, normal spatially inhomogeneous elastic solids would couple compression and shear waves \cite{Stenger12,Milton02,Milton06,Norris08,Norris09,Farhat09}. Thus, an incident compression wave would lead to emerging shear waves as well, making the realization of transformation-elastodynamic structures in three dimensions impossible. Pentamodes are a mathematical ideal \cite{Milton02,Milton95} for which the shear modulus is strictly zero, thus no shear modes exist at finite frequencies. Recently, we have fabricated 3D microstructures \cite{Kadic12} that approach the pentamode ideal in that the effective bulk modulus computed within static continuum mechanics is more than three orders of magnitude larger than the effective shear modulus. 

However, it is not obvious how these static results can be translated to the dynamic case, i.e., to phonon waves. For example, one would like to know the phase velocities of the different modes. The moduli are known from static continuum mechanics. For obtaining knowledge on phase velocities, one additionally needs the effective mass densities. It is known that the dynamic effective mass density can be distinct from the static mass density as one would determine by using a mass balance (see, e.g., \cite{Berryman80,Krokhin03,Mei06}).
\begin{figure}[h!]
	\centering
	\includegraphics[width=0.4\textwidth]{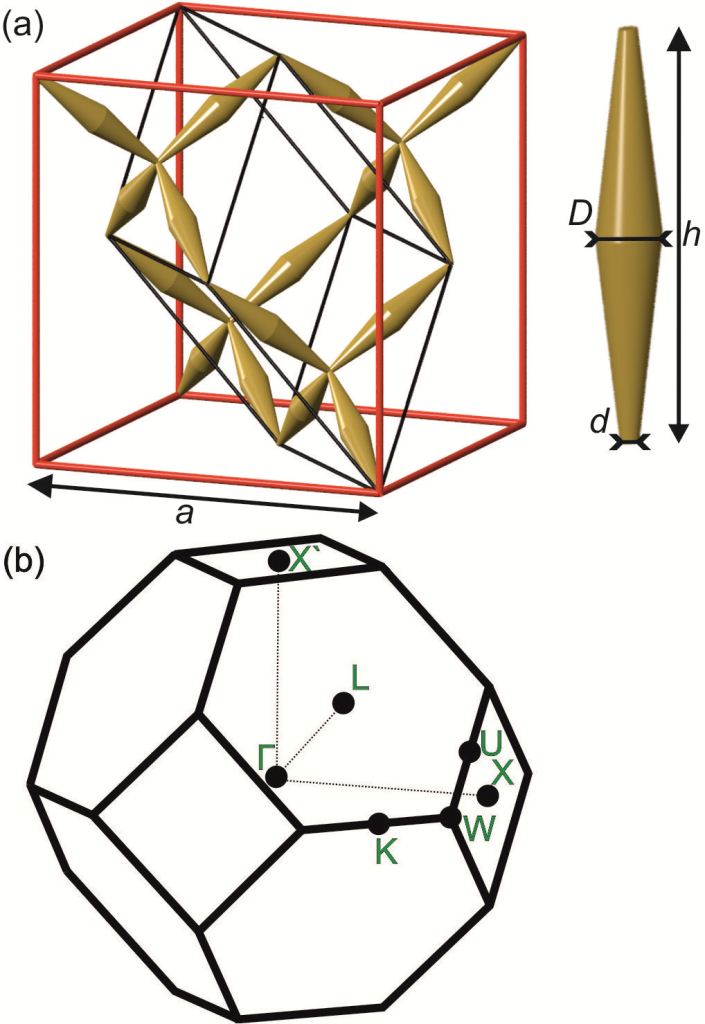}%
	\caption{(a) Illustration of the structure approximating the pentamode metamaterial ideal considered in this Letter. Both the extended face-centered-cubic (fcc) unit cell (red) and the primitive cell (black) are shown. (b) Corresponding body-centered-cubic (bcc) Brillouin zone. The characteristic points are indicated, aiming at easing the interpretation of the band structures in Fig.~2 and Fig.~4.}
	\label{fig1}
\end{figure}
Furthermore, it has not been clear up to which maximum operation frequency pentamodes would keep their desired effective material properties. Band structures have been published for honeycomb-lattice structures \cite{Martinsson03} that may be seen as the two-dimensional counterpart of pentamodes (i.e., as bimodes), but not for three-dimensional pentamodes so far.
Thus, in this Letter, we compute and discuss pentamode phonon band structures systematically as a function of the underlying structure parameters. To connect to our previous static work \cite{Kadic12} and to allow for a direct comparison, we choose parameters for the constituent material corresponding to our recent work.

\begin{figure}[h!]
	\centering
	\includegraphics[width=0.45\textwidth]{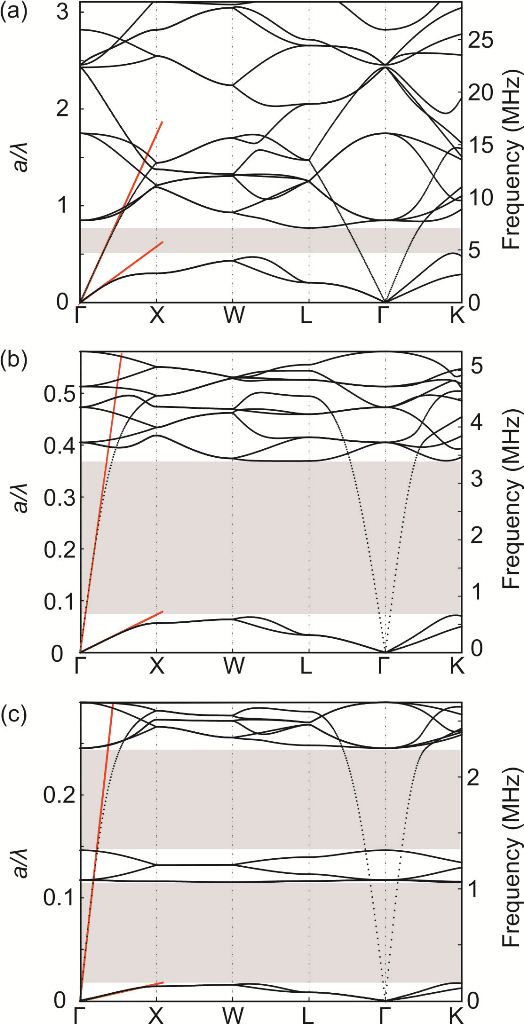}%
	\caption{Calculated band structures for the pentamodes shown in Fig.~\ref{fig1}. The fixed parameters are $a=\unit{37.3}{\micro\metre}$ (hence $h=\unit{16.15}{\micro\metre}$) and $D=\unit{3}{\micro\metre}$. The diameter of the connection region $d$ (see Fig.~\ref{fig1}) is varied from (a) $\unit{3}{\micro\metre}$ to (b) $\unit{0.55}{\micro\metre}$ to (c) $\unit{0.2}{\micro\metre}$. The red straight lines are fits to the dispersion branches of interest in this Letter. They correspond to phase velocities of (a) $c_G=\unit{198}{\metre\per\second}$ and $c_B=\unit{596}{\metre\per\second}$, (b) $c_G=\unit{25}{\metre\per\second}$ and $c_B=\unit{363}{\metre\per\second}$, and (c) $c_G=\unit{6}{\metre\per\second}$ and $c_B=\unit{240}{\metre\per\second}$. In the regions highlighted in gray, only a single phonon mode occurs due to a complete three-dimensional band gap for shear waves. By the scalability of the problem one can, for example, simultaneously replace \micro\metre\ by \milli\metre\ and \mega\hertz\ by \kilo\hertz.}
	\label{fig2}
\end{figure}

Fig.~\ref{fig1}(a) recalls the structure approximating the pentamode ideal suggested by Milton and Cherkaev \cite{Milton95}. Both the extended face-centered-cubic (fcc) cell with lattice constant $a$ and the primitive cell (black) are depicted. Cones touch each other at their thin ends with diameter $d$. These connection points form a diamond lattice. Pairs of cones are connected at their thick ends with diameter $D$. The length of the double-cones is then given by $h=\sqrt{3}\,a/4$. Panel (b) of Fig.~\ref{fig1} illustrates the corresponding body-centered-cubic (bcc) Brillouin zone (i.e., that of diamond). The usual characteristic points are indicated, aiming at easing the interpretation of the corresponding band structures shown in Fig.~\ref{fig2} and Fig.~\ref{fig5}. 

For the numerical calculation of the pentamode band structure, we choose a polymer as constituent material with Young’s modulus \unit{3}{\giga\pascal}, Poisson’s ratio 0.4, and mass density $\rho = \unit{1190}{\kilo\gram\per\cubic\metre}$ (as previously \cite{Kadic12}). We solve the elastodynamic equations \cite{Milton02} in vacuum. The results should also be applicable to pentamodes in air as long as the air speed of sound does not match that of the pentamode branches, in which case strong coupling could occur. The coupled elastodynamics-acoustics problem has only been addressed in very rare cases \cite{Estrada09}. The scalability of the elastodynamic equations allows to easily translate our results to other choices of the Young’s modulus, provided the Poisson’s ratio as well as the structure parameters are fixed. For example, if one increases the Young’s modulus by factor 100, the frequency increases by the square root, i.e., by factor 10. We use a commercial software package (\textsc{Comsol} Multiphysics, \textsc{Mumps} solver) with Bloch boundary conditions imposed for the primitive real-space cell shown in Fig.~\ref{fig1}(a). We have carefully checked that all results depicted are converged. Typically, convergence is achieved by using around $4\times10^4$   tetrahedra in one primitive real-space cell. Also, we choose $a=\unit{37.3}{\micro\metre}$ throughout as previously \cite{Kadic12}. The right-hand side vertical frequency scale in Fig.~\ref{fig2} is in S.I.\ units. On the left-hand side vertical scale, we also show the $a/\lambda$ ratio as a normalized and more universal frequency scale. Here, $\lambda$ is the wavelength of sound in air, for which we have taken a standard sound velocity of $\unit{343}{\metre\per\second}$. For example, a ratio of $a/\lambda=0.1$ means that the air wavelength is ten times larger than the lattice constant $a$ (whatever the value of $a$ might be). The usual tour through the Brillouin zone on the horizontal axis is parameter-independent anyway.

In Fig.~\ref{fig2}, we fix $D=\unit{3}{\micro\metre}$ and vary $d$ from (a)
\unit{3}{\micro\metre} to (b) \unit{0.55}{\micro\metre} to (c)
\unit{0.2}{\micro\metre} (compare insets in Fig.~\ref{fig4}). Note the different
vertical scales. The branches emerging from the $\Gamma$ point (i.e., wave
vector $= (0,0,0)$, see Fig.~\ref{fig1}(b)) are the dispersions of the
“acoustic” shear and compression waves respectively. Their slopes  are the
corresponding phase velocities (see red straight lines). Obviously, the ratio of the slopes increases
with decreasing diameter $d$. This trend was to be expected from the static
continuum-mechanics calculations in Ref.~\cite{Kadic12}. A large ratio of the
slopes is desirable for transformation elastodynamics, because it suppresses
undesired coupling between shear and compression waves in spatially
inhomogeneous architectures. We will come back to the dependence of this ratio
below. Furthermore, the gray regions in Fig.~\ref{fig2} highlight frequency
intervals in which only a single phonon mode exists in the pentamode structure.
The relative width of the lower frequency interval increases with decreasing diameter $d$ from (a) to (c) in Fig.~\ref{fig2}. Our previous experiments \cite{Kadic12} have realized about $d=\unit{0.55}{\micro\metre}$. In the corresponding Fig.~\ref{fig2}(b), the upper frequency bound of this interval is more than two octaves larger than its lower bound. In these single-mode regimes, only compression waves exist and shear waves are suppressed by a complete three-dimensional band gap. We will speculate about possible applications of this finding at the end of this Letter. At yet higher frequencies, we find the usual “spaghetti” region.

\begin{figure}
	\centering
	\includegraphics[scale=0.85]{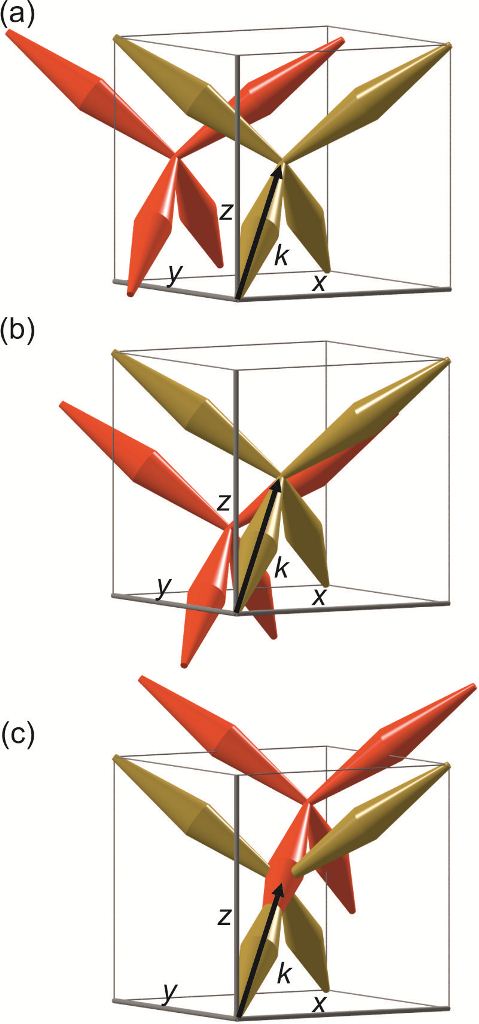}
	\caption{Illustration of selected modes for parameters corresponding to Fig.~\ref{fig2}(b). The wave vector $\vec{k}$ is along the $\Gamma\text{K}\parallel(1,1,1)$ direction (see arrow). Its modulus is small compared to $\pi/a$. In yellow we show the undisplaced structure (compare Fig.~\ref{fig1}(a)) and in red a snapshot of the displaced structure (largely exaggerated). (a), (b), and (c) correspond to the first, second, and third band. The displacement vectors, normalized to unity length, are: (a) $\vec{u}=1/2\,(-1,\sqrt{3},0)$, (b) $\vec{u}=1/2\,(0,\sqrt{3},-1)$, and (c) $\vec{u}=1/\sqrt{3}\,(1,1,1)$.}
	\label{fig3}
\end{figure}

Examples of mode shapes are depicted in Fig.~\ref{fig3}. Here we choose parameters as in Fig.~\ref{fig2}(b) and propagation along the $\Gamma$K direction, which corresponds to the space diagonal of the extended fcc cell shown in Fig.~\ref{fig1}. The results shown refer to the limit of a small wave number. For larger wave numbers, visualization of the modes requires depicting several or many unit cells. Panels (a), (b), and (c) in Fig.~\ref{fig3} refer to the first, second, and third band. Clearly, panel (c) exhibits a compression wave with a displacement of the double cones along the wave vector (see arrow), whereas (a) and (b) show shear waves, for which the displacements contain components orthogonal to the wave vector.

To get an intuitive understanding for the band structures, let us consider the expectation for the pentamode ideal. Here, the shear modes have strictly zero frequency and, in addition to the compression-mode dispersion with finite slope, one gets different strictly localized vibrations of the isolated double cones (which only touch each other at singular points in the pentamode ideal, see Fig.~\ref{fig1}). These localized vibrations in real space correspond to flat bands in momentum space. The behavior shown in Fig.~\ref{fig2}(c) obviously approaches that ideal. Here, the finite frequency extent of the fairly flat bands results from the finite (i.e., not quite point-like) connections with diameter $d$ between the double cones. This reasoning also qualitatively explains the occurrence of the single-mode regions in between the spatially fairly localized modes.

 \begin{figure}[h!]
 	\centering
 \includegraphics[width=0.45\textwidth]{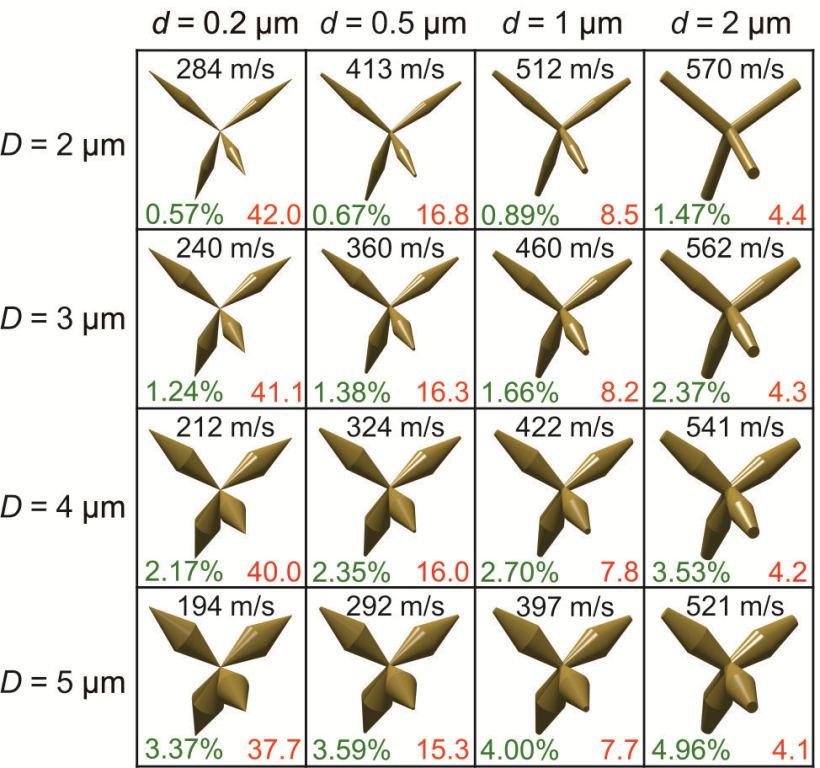}%
 \caption{The ratio $c_B/c_G$ (red) of the phase velocities of compression (black) and shear waves (compare red straight lines in Fig.~\ref{fig2}) for different parameter combinations of $d$ and $D$ (see Fig.~\ref{fig1}) as derived from band structure calculations like shown in Fig.~\ref{fig2} for the $\Gamma \text{X}$ direction (i.e., propagation of the waves along any of the six face diagonals of the cube in Fig \ref{fig1}(a)). The volume filling fractions $f$ are given in green.}
 \label{fig4}
 \end{figure}

The dependence on the large diameter $D$ is much less pronounced. To give an overview, the matrix structure in Fig.~\ref{fig4} shows the ratio of the phase velocities of compression and shear waves for various combinations of $D$ and $d$. In each box of this matrix, the structure is illustrated too. The phase-velocity ratio is nearly independent on $D$ and scales roughly inversely with $d$. This qualitative scaling is consistent with our previous static continuum-mechanics calculations \cite{Kadic12}. There, we found that the ratio of the bulk modulus $B$ and the shear modulus $G$ scales approximately like
\begin{equation*}
\frac{B}{G}\approx 0.63 \left(\frac{h}{d}\right)^2 \sqrt{\frac{h}{D}} \quad \text{.}
\end{equation*}
For isotropic media in continuum mechanics, the phase velocity of the compression wave, $c_B$,  is given by
\begin{equation*}
  c_B=\sqrt{\frac{B+4\,G/3}{\rho_B}} \approx \sqrt{\frac{B}{\rho_B}} \quad \text{,}
\end{equation*}
with the effective (dynamic) mass density $\rho_B$. In the last step, we have assumed $B/G \gg 1$. For the shear waves, one has 
\begin{equation*}
  c_G=\sqrt{\frac{G}{\rho_G}} \quad \text{.}
\end{equation*} 
Naively, one would assume that both mass densities $\rho_B=\rho_G=\rho_0$  can be taken as the static mass density $\rho_0$, which is what one would measure with a mass balance in the laboratory.  Mathematically, $\rho_0=f\rho$ is simply given by the volume filling fraction $f$ times the mass density $\rho$ of the constituent material. The more sophisticated Berryman formula \cite{Berryman80} is often a more accurate approach but is also known to fail quantitatively and qualitatively in some cases \cite{Krokhin03}. In any case, when taking the ratio $c_B/c_G$, any common mass density would drop out and one gets a scaling according to $c_B/c_G \propto 1/(d D^{1/4})$ $\propto~1/d$. Indeed, we obtain a good least-squares fit to the numerically computed values in Fig.~\ref{fig4} with the formula
\begin{equation*}
  \frac{c_B}{c_G} \approx 0.5\, \frac{h}{d} \quad \text{.}
\end{equation*}
For example, for $d=\unit{0.55}{\micro\metre}$, $D=\unit{3}{\micro\metre}$, and $a=\unit{37.3}{\micro\metre}$, hence $h=\unit{16.15}{\micro\metre}$, we have $c_B/c_G \approx 16$ in Fig.~\ref{fig4}. For the same parameters, the above heuristic continuum-mechanics formula yields $B/G \approx 1260$ \cite{Kadic12}, which, assuming identical mass densities $\rho_B=\rho_G$, leads to $c_B/c_G \approx \sqrt{1260} \approx 35$. Considering the simplicity of this reasoning, the result is not too far off.  One should also be aware that there is no unique phase velocity of the shear waves as can be seen from the splitting of the corresponding dispersion branches in the $\Gamma \text{K}$ direction in Fig.~\ref{fig2}.  We will come back to the aspect of isotropy below.

In static continuum mechanics \cite{Kadic12}, we found very little dependence of the results on the Poisson’s ratio of the constituent material (the polymer). We have also investigated this aspect for the band structures. For example, when drastically changing the Poisson’s ratio from 0.4 (as above) to 0.1 and keeping all other parameters fixed, the relevant frequencies in the band structure shift by less than \unit{3}{\%} (not depicted). However, changing the Young’s modulus of the constituent material will, of course, stretch/compress the frequency scale by a common factor as discussed above but it will not change the qualitative behavior of the band structure at all. Likewise, a change in lattice constant $a$ leads to a trivial inverse scaling of the frequency axis.

 \begin{figure}[h!]
 	\centering
 \includegraphics[width=0.45\textwidth]{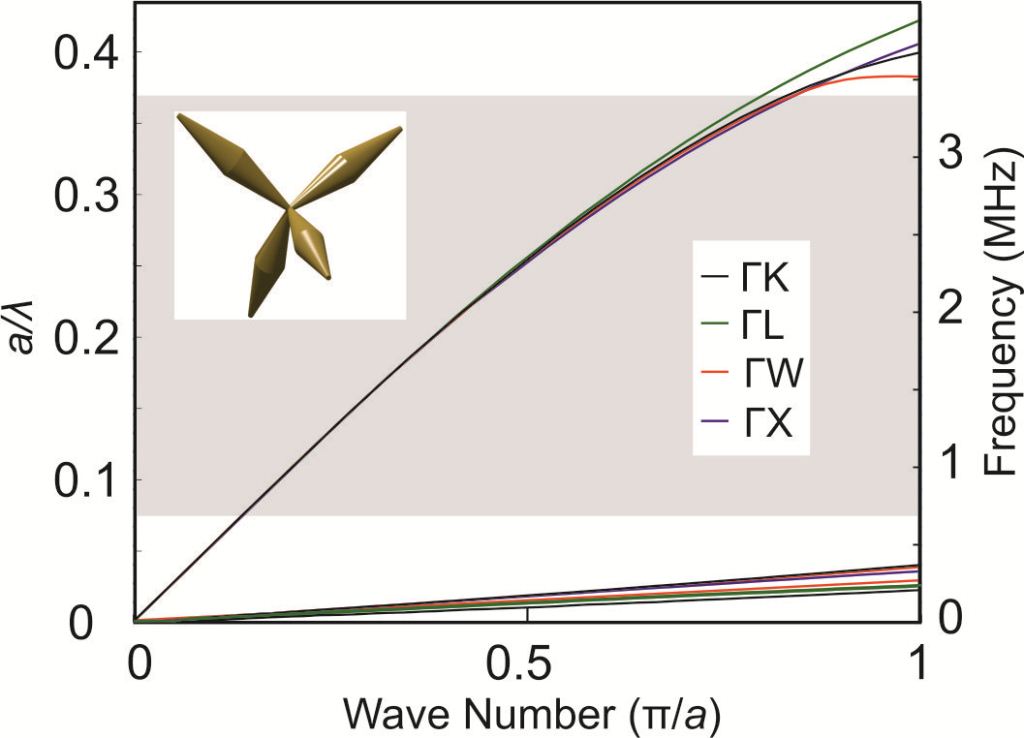}%
 \caption{Pentamode dispersion relations for four selected characteristic directions in the first Brillouin zone (compare Fig.~\ref{fig1}(b)). The parameters and the gray single-mode region are as in Fig.~\ref{fig2}(b). As in Fig.~\ref{fig2}(b), we obtain four compression-wave branches and, due to the lifting of degeneracy in the $\Gamma$K and $\Gamma$W directions, six shear-wave branches.}
 \label{fig5}
 \end{figure}
 
Let us come back to the above mentioned single-mode regime. In Fig.~\ref{fig2}(a) for $d=\unit{0.55}{\micro\metre}$, its upper limit lies at a normalized frequency of about $a/\lambda=0.4$ (the lower limit lies below $a/\lambda=0.1$). This means that the corresponding wavelength in air is 2.5 times larger than the pentamode fcc lattice constant $a$ at the upper limit. Below this upper frequency, the dispersion can still be approximated reasonably well by a straight line. One might expect that the effective medium approximation is still reasonably well fulfilled here too. Also, we would still expect a more or less isotropic behavior of the compression waves – as for an ideal pentamode \cite{Milton95}. To investigate this aspect, we depict in Fig.~\ref{fig5} the numerically calculated dispersion relations for various directions in momentum space for $d=\unit{0.55}{\micro\metre}$ (i.e., all parameters as in Fig.~\ref{fig2}(b)). Obviously, the steeper compression-wave branches in Fig.~\ref{fig5} exhibit only very little dependence on direction, i.e., the behavior is very nearly isotropic in 3D. In contrast, differences in the slopes as large as a factor of two occur for the flatter shear-wave branches. Indeed, it was already apparent from the lifting of degeneracy (two branches instead of one) in the dispersions in the $\Gamma \text{K}$ and the WL directions in Fig.~\ref{fig2} that the pentamode shear-wave dispersion is not isotropic at all (also see Fig.~\ref{fig3}). This is consistent with our previous treatment using static continuum mechanics, where the shear modulus also depends on direction. In \cite{Kadic12} we have only shown the shear modulus for shear along an edge of the extended fcc unit cell.

Finally, we speculate that the single-mode region (Fig.~\ref{fig2}) with nearly isotropic behavior for the compression waves and with a complete three-dimensional band gap for shear waves might be interesting for future three-dimensional transformation-elastodynamic architectures. After all, one motivation for constructing pentamode metamaterials lies in suppressing the shear waves in 3D. One should be aware though that the different local single-mode regions of a generally complex, spatially inhomogeneous and possibly also anisotropic pentamode-based elastodynamic architecture will likely not fully overlap. However, one could choose the operation frequency to be within the single-mode region in the outer parts of the overall structure. In this case, at least no shear waves could emerge, e.g., from a cloak.

In conclusion, we have calculated phonon band structures for 3D metamaterials approximating the pentamode ideal for the first time. Our findings agree qualitatively with our previous static continuum-mechanics calculations in that we find ratios of the phase velocities of compression and shear waves as large as 16 for experimentally accessible structure parameters. Moreover, we also find an interesting single-mode frequency interval with a width larger than two octaves. For reasonable material parameters, its upper frequency bound corresponds to an air wavelength that is only 2.5 times larger than the pentamode fcc lattice constant. This means, for example, that order $a=\unit{1}{\centi\metre}$ pentamode lattice constants would suffice for up to \unit{10}{\kilo\hertz} operation frequencies. Within this single-mode regime, the compression-wave dispersion is still nearly isotropic in three dimensions.

We thank Andreas Frölich, Jonathan Müller, and Michael Thiel for help with implementing the periodic boundary conditions. We acknowledge support by the DFG-Center for Functional Nanostructures (CFN) via subproject A1.5.



\providecommand{\noopsort}[1]{}\providecommand{\singleletter}[1]{#1}%

\end{document}